\documentclass[aps,prl,twocolumn,groupedaddress]{revtex4}

\usepackage{graphicx}

\begin{document}


\title{Sliding conduction by the quasi one-dimensional charge-ordered state in Sr$_{14-x}$Ca$_x$Cu$_{24}$O$_{41}$}


\author{A. Maeda}
\email[E-mail\ \ ]{maeda@maildbs.c.u-tokyo.ac.jp}
\homepage[Web\ \ ]{http://maeda3.c.u-tokyo.ac.jp}
\author{R. Inoue}
\author{H. Kitano}
\affiliation{Department of Basic Science, University of Tokyo\\
                3-8-1, Komaba, Meguro-ku, Tokyo, 153-8902, JAPAN}
\author{N. Motoyama}
\author{H. Eisaki}
\author{S. Uchida}
\affiliation{Department of Advanced Materials Science, University of Tokyo\\
             7-3-1, Hongo, Bunkyo-ku, Tokyo, 113-8586, JAPAN}


\date{Received June 5, 2002}

\begin{abstract}

Nonlinear conduction (NLC) of the two-leg spin ladder, Sr$_{14-x}$Ca$_x$Cu$_{24}$O$_{41}$, was investigated for the $x$=0, 1 and 12 materials .  Although insulating materials ($x$=0 and 1) exibited the NLC both in the ladder- and rung directions, the NLC in the ladder direction of the $x$=0-material was found to be very special.  We considered this to be due to the sliding motion of the charge ordered state, which was responsible for the resonance at microwave frequencies.
We discussed possible candidates for the charge ordered state responsible for the NLC, including Wigner crystal in quasi one dimension (4$k_F$-CDW).

\end{abstract}


\maketitle


Sr$_{14-x}$Ca$_x$Cu$_{24}$O$_{41}$ (SCCO) has two kinds of quasi one-dimensional structures constructed from Cu and O\cite{Carter}.
One is the CuO$_2$ chain and the other is the Cu$_2$O$_3$ ladder, which can be regarded as the one-dimensional offshoot
of the two-dimensional CuO$_2$ planes.
Thus, this material has been thought to be a good reference for high-$T_c$ cuprate superconductors, where the square CuO$_2$ plane is essential for high-$T_c$ superconductivity.
Dagotto and Rice\cite{Dagotto} predicted that carrier doping into such a ladder structure of a strongly correlated system leads to superconductivity, through a mechanism which they thought was similar to that of high-$T_c$ superconductivity in quasi two-dimensional cuprates.
In fact, superconductivity was realized under high pressure in a material heavily doped by Ca\cite{Akimitsu}.
According to an optical study\cite{Osafune}, the role of Ca substitution, which does not change the formal valence of Cu, was thought to transfer holes from the chain to the ladder.
Thus, the superconductivity in this system is considered to be realized when the quasi one-dimensionality of holes on the ladder becomes sufficiently weakened\cite{Nagata}.

The opposite limiting case of low hole concentration is also interesting.
The c-axis (ladder-direction) dc conductivity of the parent material, Sr$_{14}$Cu$_{24}$O$_{41}$ (SCO), is insulating with an activation gap of $\sim$ 2000 K\cite{Motoyama} at least below 300 K.  An NMR/NQR measurement suggested the occurrence of some charge order below 150 K\cite{Takigawa}.
A recent study of the microwave electrical conductivity found a dramatic increase of conductivity below $\sim$ 170 K\cite{Kitano}.  Detailed study of the frequency dependence of the conductivity showed a small but sharp resonance around 50 GHz ($\sim$ 2.4 K) in this temperature region.
Since the peak frequency is much smaller than the thermal energy per carrier, this strongly suggested the resonance of some collective charge excitation, similar to the pinned phase mode of charge-density wave (CDW) systems and spin-density wave (SDW) system in quasi-one dimensional materials\cite{Gruner}.
It also reported a definite nonlinear electrical conduction as a function of electric field in a rather weak field region.
Quite recently, a similar NLC was also observed by Gorshunov {\it et al}\cite{Gorshunov}.
This NLC also suggested the collective nature of the phenomena.
These phenomena are quite reminiscent of those established in the CDW systems.
However, a very small oscillator strength of the collective resonance in SCO showed a serious discrepancy in quantitative aspects of the interpretation as being a resonance of the pinned CDW.  Thus, it has been suggested that the collective mode found in SCO might be a mode characteristic of a strongly correlated system.

To investigate the details of this collective mode, we have studied the nonlinear conduction as a function of dc electric field at various temperatures, in samples with different carrier concentrations, and along different directions of the crystal.
We found the nonlinear conduction to be present in almost all experiments (samples, temperatures, directions).  However, the nonlinear conduction found in the c-direction of SCO was very different from all other results.  Thus, we conclude that the collective mode has a quasi one-dimensional nature, and is characteristic of samples with low hole concentration.
We discussed possible candidates for the charge ordered state responsible for the NLC, including Wigner crystal (WC)\cite{Wigner} in quasi one dimension (4$k_F$-CDW).


Crystals were prepared by the traveling-solvent-floating-zone method\cite{Motoyama}.
For electrical conductivity measurement, small pieces were cut from the boul, with typical dimensions of 3 $\times$ 1 $\times$ 0.5 mm$^3$.  The longest dimension was in the c-axis.
Electrical contacts were made by evaporating silver on the sample, above which electrical leads were attached by EPOTEK-H20E silver paste.
Post annealing under an oxygen flow atmosphere at 200 degrees centigrade decreased the contact resistance remarkably.
Typical contact resistance was 10$^4$ $\Omega$ at 100 K ($x$=0), which was about 20 \% of the intrinsic resistance of the sample.
So, it is important to check the ohmic nature of the electrical contact, which was confirmed in each measurement.
The electrical conductivity was measured by four-probe methods.
To avoid the temperature rise due to Joule heating, short rectangular pulses generated by an HP8116A pulse function generator, amplified by an NF4105 amplifier, were applied to the sample, and the response was measured by an SRS-SR250 boxcar averager or a TDS420A digital oscilloscope.
For both cases, an NF5305 differential preamplifier was used if necessary.
By repeating the measurements for pulses with different repetition rates (10$\sim$1 kHz), duty cycles (1/10 $\sim$ 1/1000), and widths (5$\sim$ 500 $\mu$sec), we checked whether or not any changes in conductivity were due to the simple Joule heating effects.
At weak electric fields, a conventional lock-in technique was also used for the differential conductivity measurement.
For measurements at various temperatures, an Oxford He-flow type cryostat was used.
To avoid the direct contact of the sample with flowing gas, we fabricated the modified insert.
By using this insert, temperature stability of about 50 mK was achieved throughout the measurements.

\begin{figure}
\includegraphics[width=7cm]{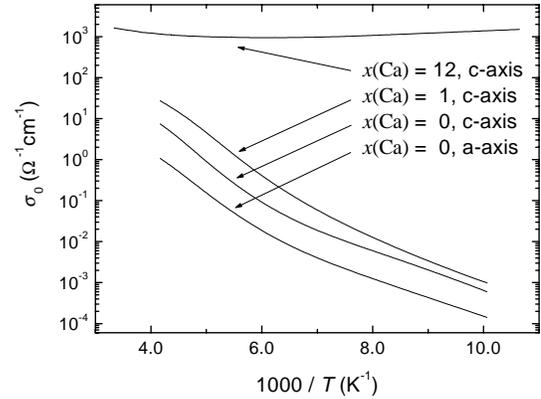}
\caption{\label{condtemp}
Low field ohmic dc conductivity, $\sigma_0$, as a function of temperature for $x$=0, 1, and 12 samples.}
\end{figure}
\begin{figure}
\includegraphics[width=7cm]{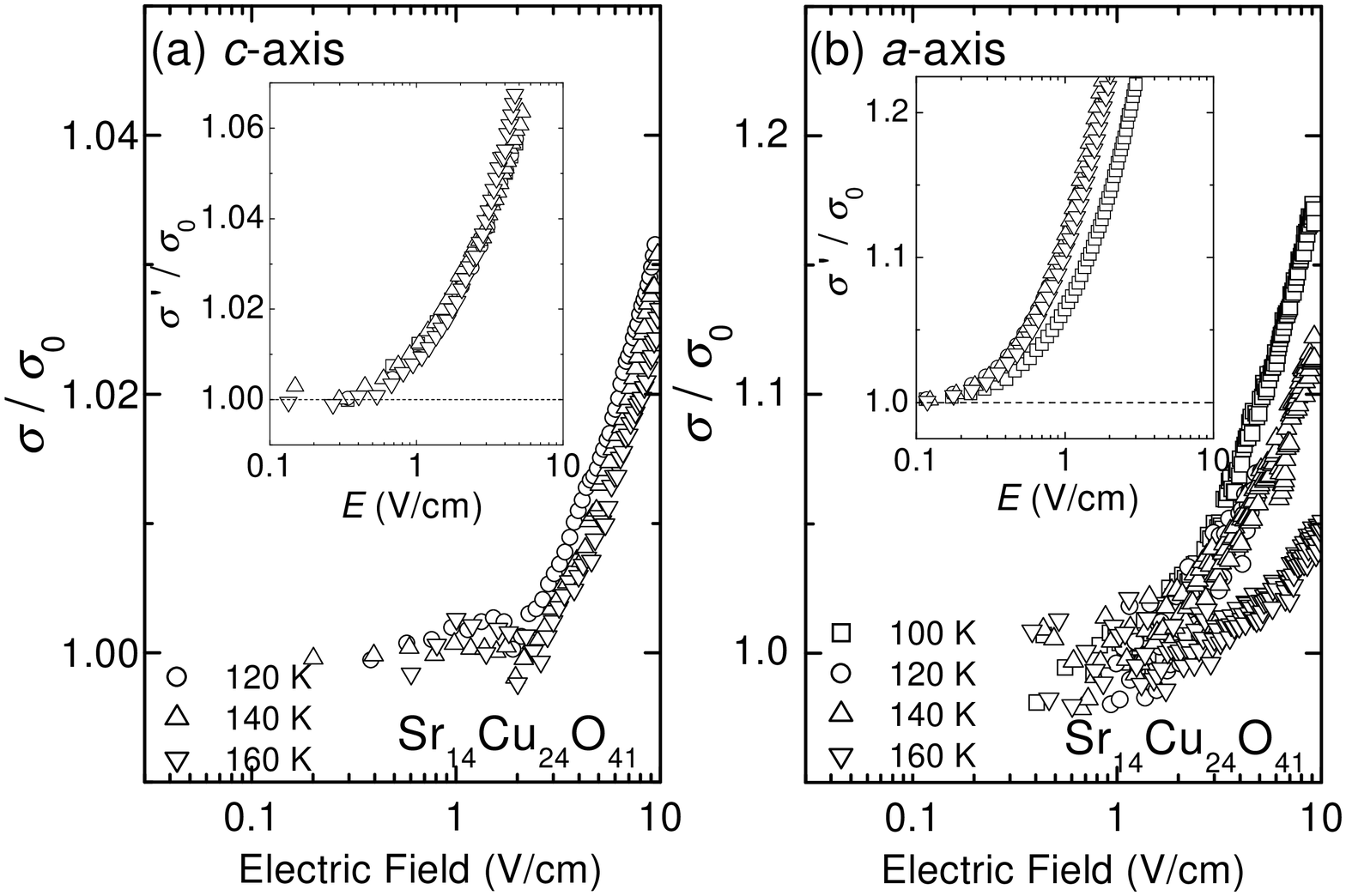}%
\caption{\label{x0condelec}
Normalized chordal conductivity of the $x$=0 material as a function of electric field at various temperatures. (a) c (ladder) direction (b) a (rung) direction.
The inset shows the differential conductivity, $\sigma'\equiv \frac{{\rm d}j}{{\rm d}E}$ at the corresponding temperatures.  Horizontal dashed lines are guides for eye to show the ohms's law.}
\end{figure}

Figure \ref{condtemp} shows the low-field conductivity, $\sigma_0$, as a function of inverse temperature for several different carrier concentrations.
Except for the $x$=12 sample, the materials are insulating with similar activation energies of $\sim$ 2500 K and 1400 K above and below 160 K, respectively.
Figure \ref{x0condelec} shows the chordal conductivity, $\sigma$, of SCO in the two different directions (ladder(c-axis) and rung(a-axis)) as a function of electric field for various temperatures.
In the inset, the differential conductivity, $\sigma'\equiv \frac{{\rm d}j}{{\rm d}E}$, is also shown for comparison.
All the data in this figure were normalized by the ohmic conductivity, $\sigma_0$, at each temperature.
In both directions, we observed nonlinearity in the conductivity as a function of electric field.
As we already reported previously, the c-axis conductivity increases with increasing field.
However, the amount of the conductivity increase is very small (at most 3 \% of the ohmic value at 10 V/cm).
On the other hand, in the a-direction (rung direction), the conductivity increase is rather distinct.
At the lowest temperature measured ($\sim$ 100 K), the conductivity increase is more than 17 \% of the low field value.

\begin{figure}
\includegraphics[width=7cm]{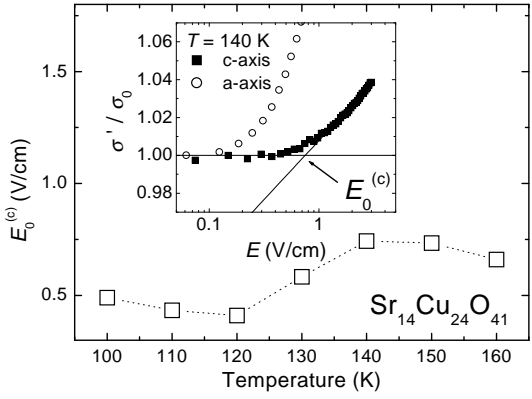}%
\caption{\label{thresh}
The characteristic field, $E_0$, for the NLC onset as a function of temperature.
Dashed lines are guides for eye.
The inset shows how $E_0$ was derived, where the a-axis data were also shown.}
\end{figure}

First, we concentrate on the c-axis conductivity.
Although it is hard to define a definite threshold field for the onset of nonlinearity, the nonlinearity seems to start between 0.1 V/cm and 2 V/cm, as shown in the inset of Fig.~\ref{thresh}.
To be quantitative, we define a characteristic field, $E_0$, for the onset of NLC as the intersection of two extrapolated straight lines in the differential conductivity data, as shown in the inset of Fig.~\ref{thresh}.
In the main panel of Fig.~\ref{thresh}, $E_0$ was displayed as a function of temperature.
$E_0$ was found not to depend very greatly on temperature.
The conductivity normalized by the ohmic value at each temperature, $\sigma_0(T)$, was found not to depend on temperature below 150 K, either.
This is remarkable because the ohmic conductivity, $\sigma_0$, changed about two orders of magnitude from 120 K to 160 K, as was already shown in Fig.~\ref{condtemp}.
This means that the conductivity in this temperature region was expressed by a formula,
\begin{equation}
\frac{\sigma(E,T)}{\sigma_0(T)} = f(\frac{E}{E_0}),
\label{empscale}
\end{equation}
where $f(z)$ is a universal function of $z$.
That is, a scaling relation holds between the nonlinear conductivity, $\sigma(E,T)$, and the ohmic conductivity, $\sigma_0(T)$.
At higher temperatures of 200 K, the Joule-heating effect hindered detailed studies of the nonlinearity.
However, it seems that the increase of the normalized conductivity becomes lessened.

On the other hand, for the a-axis nonlinear conductivity, the quantitative behavior is rather different.
First, in the $a$ direction, as shown in the inset of Fig.~\ref{thresh}, the existence of the characteristic field, $E_0$, for the onset of nonlinear conduction is less clear.
Within our resolution, the nonlinearity exists even at the lowest field.
Second, there is no temperature region where the scaling relation holds.
Such quantitative differences between the c-axis and a-axis suggest that the mechanism of the NLC is different between these two directions.

\begin{figure}
\includegraphics[width=7cm]{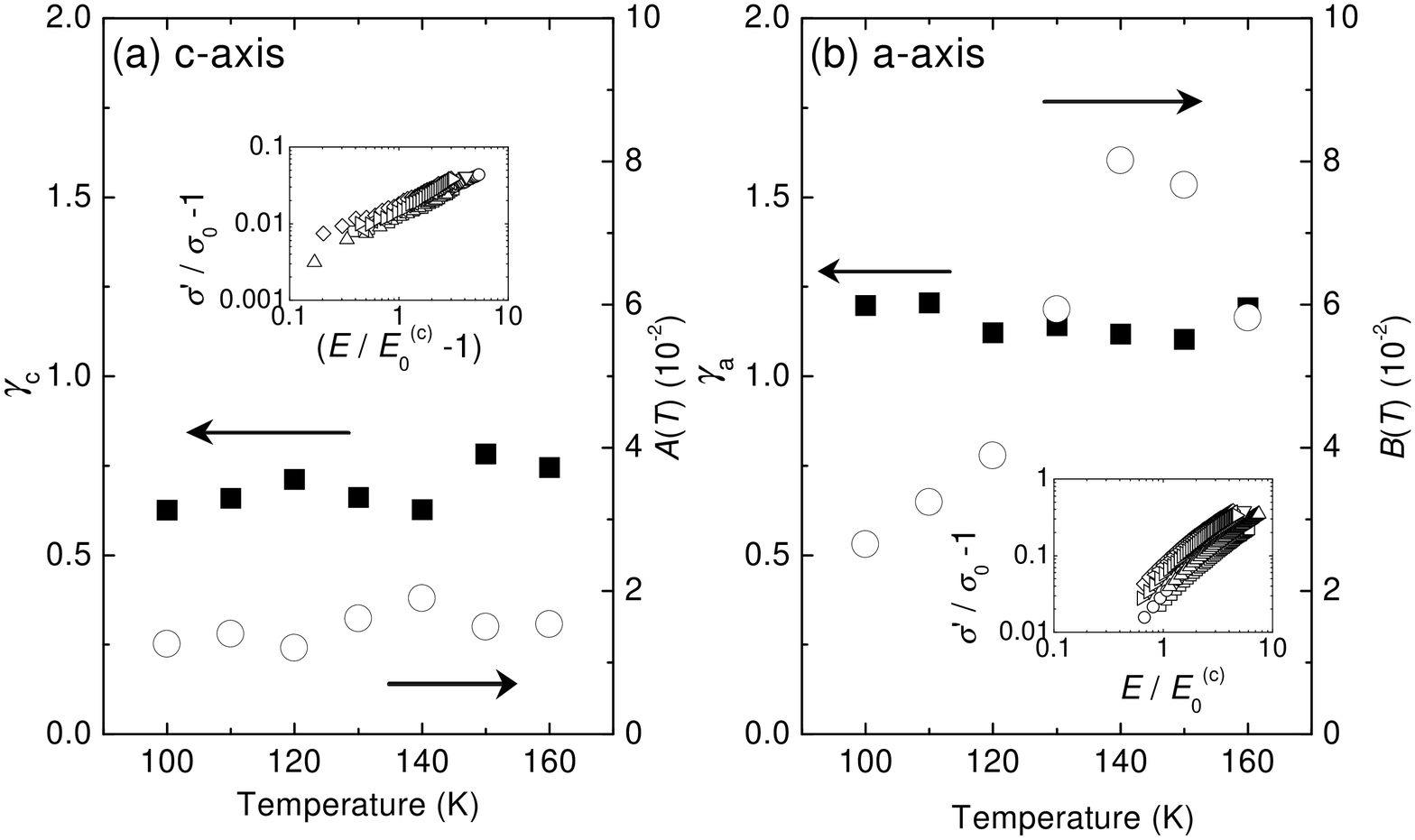}%
\caption{\label{scale}
Exponents and the proportional constants appeared in eqs. \ref{powerc} and \ref{powera}.  The inset shows the raw condeuctivity data fitted by these equations.}
\end{figure}

For further comparison, we empirically express the differential-conductivity data using power laws as follows, since the nonlinearity is not so strong.
\begin{equation}
\frac{\sigma'(E)}{\sigma_0}-1 = A(T) (\frac{E}{E_0^{(c)}}-1)^{\gamma_c} \ \ \ \ ({\rm c \ axis}),
\label{powerc}
\end{equation}  
and 
\begin{equation}
\frac{\sigma'(E)}{\sigma_0}-1 = B(T) (\frac{E}{E_0^{(c)}})^{\gamma_a} \ \ \ \ ({\rm a \ axis}),
\label{powera}
\end{equation}
where $E_0^{(c)}$ is the characteristic field for the onset of nonlinearity along the c-axis, $\sigma_0$ is the conductivity at the lowest field in each direction, $\gamma_c$, $\gamma_a$ are exponents, $A(T)$ is an almost temperature-independent constant, and $B(T)$ is a temperature-dependent constant.
If we fit the data to these formulae, as is shown in the inset of Fig.~\ref{scale},
we obtain the temperature dependence of the exponents, $\gamma_c$ and $\gamma_a$, as shown in Fig.~\ref{scale}~(a), and those of $A(T)$ and $B(T)$, as shown in Fig.~\ref{scale}~(b).
As was anticipated, $A(T)$ and $\gamma_a$, and also $\gamma_a$, did not depend on temperature, whereas $B(T)$ depended on temperature strongly.

\begin{figure}
\includegraphics[width=7cm]{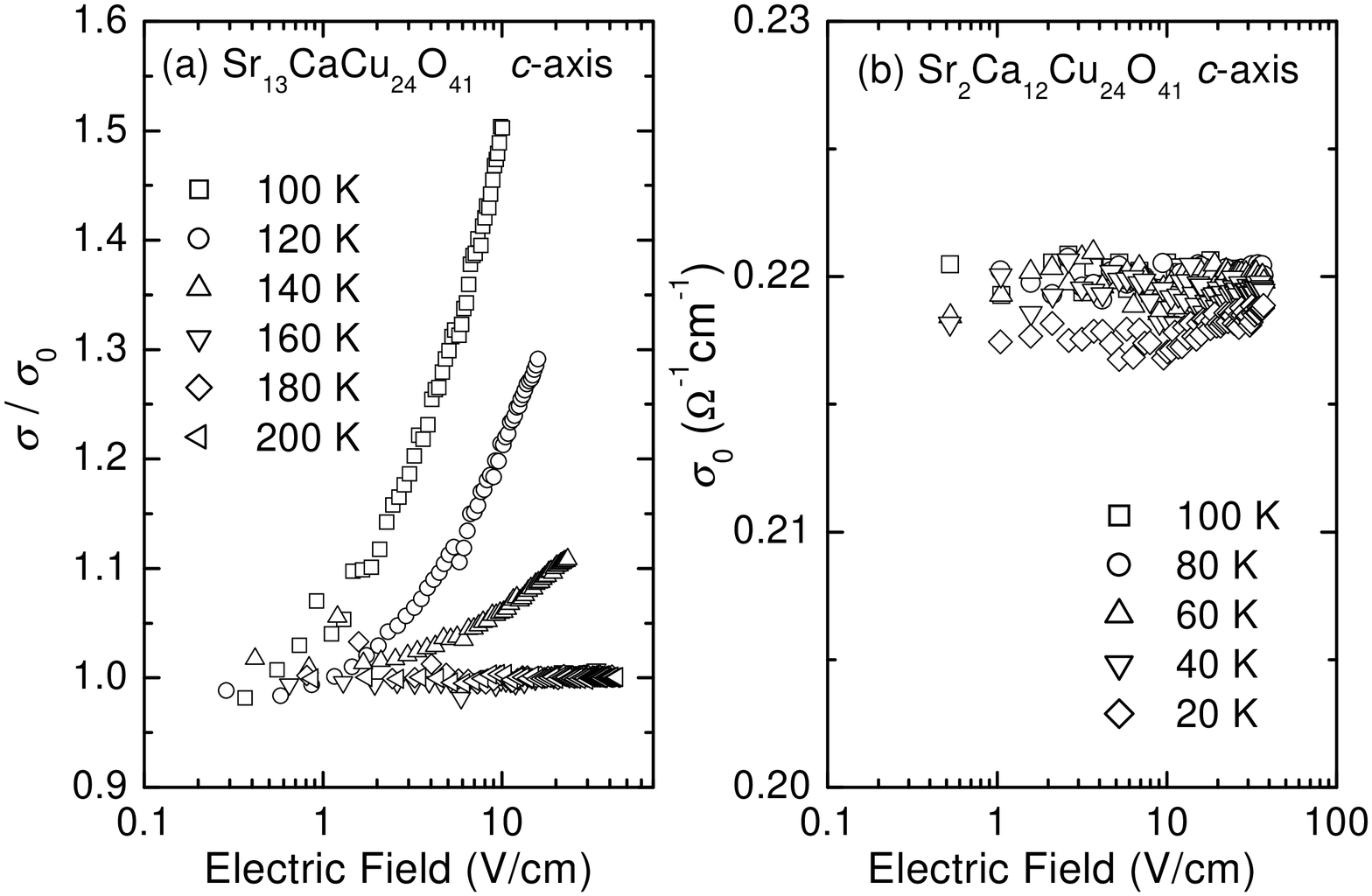}%
\caption{\label{condcomp}
C-axis chordal conductivity as a function of electric field in the $x$=1 and $x$=12 materials.}
\end{figure}

Figure~\ref{condcomp} shows the conductivity in the c-direction (ladder direction) as a function of electric field for different Ca concentrations, $x$=1 and $x$=12.
For the $x$=1 sample, the basic features of the NLC are very similar to those in the a-direction of the $x$=0 sample.
That is, the conductivity increase is large (more than 15$\sim$20 \% at 100 K) without any definite threshold nor characteristic fields for the NLC onset, and there was no sign of a scaling relation such as eqs.~\ref{empscale} and \ref{powerc}.
On the other hand, for the $x$=12 sample, the conductivity remained field-independent up to 30 V/cm within a resolution of 0.3 \%.


The above results showed that the NLC was observed only in materials with a semiconducting (or insulating) ground state.
Furthermore, the NLC in the ladder direction of the $x=$0 material showed special behavior when compared with the data in the rung direction and with those from samples with different Ca contents.
In the previous paper\cite{Kitano}, we suggested that the NLC is due to the collective motion of some charge ordered state that also exhibited a sharp resonance in the microwave region.
For the NLC due to the collective motion of the charge ordered state such as CDW, if the weak field conductivity shows semiconducting temperature dependence, it has been known\cite{Gruner} that the extra conductivity due to the collective motion of the condensate, $\Delta\sigma_{col}$, is proportional to the quasiparticle conductivity (conductivity due to the uncondensed carrier), $\sigma_{qp}$,
\begin{equation}
\Delta\sigma_{col}\propto\sigma_{qp}.
\label{DWscale}
\end{equation}
This is because the deformation of the density wave (DW) is screened by the uncondensed quasiparticles.
The c-axis conductivity of the $x$=0 sample was found to obey the scaling relation (eq.~\ref{empscale}).  This is exactly what is described in eq.~\ref{DWscale}.
Thus, as far as the c-axis NLC of the $x$=0 material is concerned, this NLC is consistent with sliding conduction due to the charge ordered condensate.

On the other hand, in the a-direction, the qualitatively different behavior of the NLC suggest that the NLC in this direction is not due to the collective motion of the charge ordered state.  This suggests a strongly one-dimensional nature of the conduction due to the collective condensate.  This is, again, consistent with the sliding motion of the charge ordered state such as the CDW in the $x$=0 material.

The absence of the characteristic field, strong temperature dependence of the prefactor, $B(T)$, and the absence of the resonance in the microwave conductivity\cite{Kitano} suggest that the NLC in the $a$ direction is due to the hopping conduction of uncondensed quasiparticles between the neighboring ladders.
The power-law field dependence suggests delocalization by the electric field from the weakly localized state of the quasiparticles\cite{Yoshida,Uchida}.

Now, let us move on to the $x$=1 material.
The NLC of the $x$=1 material in the c-direction is quite reminiscent of the a-axis NLC of the $x$=0 material, and is probably caused by the delocalization of the quasiparticles by the electric field.
  This suggests that the charge ordered state showing the sliding conduction is characteristic of material  with small carrier concentration, and is very fragile for carrier doping.
Thus, it can be said that the collective mode is closely related to electron-electron correlation.

Previously\cite{Kitano}, we discussed that the characteristic field, $E_0$ satisfied a relationship expected for the depinning of the CDW,
\begin{equation}
e^*E_T \simeq m^*\omega_0^2\lambda,
\end{equation}
where $e^*$ and $m^*$ are the effective charge and the mass of the collective mode, $E_T$ is the threshold field of the NLC, and $\omega_0$ is the pinning frequency of the collective mode, provided $e^*$=$e$=1.6$\times$10$^{-19}$ C and $m^*$=$m$=9.1$\times$10$^{-31}$ kg.
This suggests that the collective mode does not contain any lattice deformations as far as the charge dynamics are concerned.
Therefore, we suggest that the collective mode responsible for the NLC in the ladder direction of the $x$=0 material is the charge-orderd state of the conduction electrons, such as the WC\cite{Wigner} in one dimension.

Formation of the WC has been considered in quantum Hall systems and He systems at an interface\cite{Fukuyama,Andrei,Shirahama1}.
Both of them are two dimensional systems, where the low density is essential to from the WC\cite{Wigner,Mott}.

However, in quasi one-dimensional systems, it has been suggested theoretically\cite{Schulz} that the long-range nature of the Coulomb interaction always leads to a very slow decay of the 4$k_F$ ($k_F$ is Fermi wave number) components of the density-density correlation function of the electrons, suggesting the stabilization of the WC (4$k_F$-CDW) across a wide range of material parameters.

As for the dynamics in the ordered state, the sliding motion of the WC was considered theoretically in one dimension\cite{Schulz} and in two dimensions\cite{Fukuyama,Giamarchi}, and experimentally only in two dimensions\cite{Goldman,YPL1,YPL2,Tsui,YPL3,Engel,Shirahama1}.
If the NLC reported here is due to the WC, this is the first example of the sliding conduction of the WC in one-dimensional systems.


Another possible candidate for the charge ordered state is the ordered state on the CuO$_2$ chain.
Indeed, the existence of the superstructure at low temperatures on the chains was reported in  neutron\cite{Matsuda} and X-ray\cite{Cox} experiments.


In conclusion, the NLC of the two-leg spin ladder, Sr$_{14-x}$Ca$_x$Cu$_{24}$O$_{41}$, was investigated for the $x$=0, 1 and 12 materials.  Although insulating materials ($x$=0 and 1) exibited the NLC both in the ladder- and rung directions, the NLC in the ladder direction of the $x$=0-material was found to be very special.  We considered this to be due to the sliding motion of the charge ordered state, which was responsible for the resonance at microwave frequencies.
We discussed possible candidates for the charge ordered state responsible for the NLC, including Wigner crystal in quasi one dimension (4$k_F$-CDW).

\begin{acknowledgments}
We thank E. Kim, B. Gorshunov, and M. Dressel for helpful discussions.
This research was partially supported by Grant-in-Aid for Scientific Research from the Ministry of the Science, Sports and Culture of Japan.
One of the authors (RI) thanks the Japan Society for the Promotion of Science for financial support.
\end{acknowledgments}


\newpage

\end{document}